\documentstyle[12pt,epsf,amsmath,axodraw,latexsym]{article}

\def\Year{\expandafter\eatPrefix\the\year}
\newcount\hours \newcount\minutes
\def\monthname{\ifcase\month\or
January\or February\or March\or April\or May\or June\or July\or
August\or September\or October\or November\or December\fi}
\def\shortmonthname{\ifcase\month\orx
Jan\or Feb\or Mar\or Apr\or May\or Jun\or Jul\or
Aug\or Sep\or Oct\or Nov\or Dec\fi}

\def\TimeStamp{\hours\the\time\divide\hours by60%
\minutes -\the\time\divide\minutes by60\multiply\minutes by60%
\advance\minutes by\the\time%
${\rm \shortmonthname}\cdot   \if\day<10{}0\fi\the\day\cdot   \the\year
\qquad\the\hours:\if\minutes<10{}0\fi\the\minutes$}

\newskip\humongous \humongous=0pt plus 1000pt minus 100pt
\def\caja{\mathsurround=0pt}
\def\eqalign#1{\,\vcenter{\openup1\jot \caja
       \ialign{\strut \hfil$\displaystyle{##}$&$
        \displaystyle{{}##}$\hfil\crcr#1\crcr}}\,}
\newif\ifdtup

\newcounter{eqnumber}[section]
\renewcommand{\theeqnumber}{\thesection.\arabic{eqnumber}}
\def\equn{\refstepcounter{eqnumber}
\eqno({\rm \theeqnumber})
}

\def\npb#1#2#3{{\rm Nucl. Phys. B}{\bf \ #1}, #3 (#2)}

\def\cqg#1#2#3{{\rm Class. and Quant.\ Grav.} {\bf  #1}, #3 (#2)}

\newbox\charbox
\newbox\slabox
\def\s#1{{      
        \setbox\charbox=\hbox{$#1$}
        \setbox\slabox=\hbox{$/$}
        \dimen\charbox=\ht\slabox
        \advance\dimen\charbox by -\dp\slabox
        \advance\dimen\charbox by -\ht\charbox
        \advance\dimen\charbox by \dp\charbox
        \divide\dimen\charbox by 2
        \raise-\dimen\charbox\hbox to \wd\charbox{\hss/\hss}
        \llap{$#1$}
}}

\def\spa#1.#2{\left\langle#1\,#2\right\rangle}
\def\spb#1.#2{\left[#1\,#2\right]}
\def\lor#1.#2{\left(#1\,#2\right)}

\catcode`@=11  

\def\Slash#1{\hskip 0.05 cm \slash\hskip -0.22 cm #1}

\def\la{\langle}
\def\ra{\rangle}

\def\lsl{\not{\hbox{\kern-2.3pt $\ell$}}}
\def\ksl{\not{\hbox{\kern-2.3pt $k$}}}

\def\spa#1.#2{\left\langle#1\,#2\right\rangle}
\def\spb#1.#2{\left[#1\,#2\right]}
\def\lor#1.#2{\left(#1\,#2\right)}

\def\sand#1.#2.#3{%
  \left\langle\smash{#1}{\vphantom1}\right|{#2}%
  \left|\smash{#3}{\vphantom1}\right\rangle}
\def\sandp#1.#2.#3{%
  \left\langle\smash{#1}{\vphantom1}^{-}\right|{#2}%
  \left|\smash{#3}{\vphantom1}^{+}\right\rangle}
\def\sandpp#1.#2.#3{%
  \left\langle\smash{#1}{\vphantom1}^{+}\right|{#2}%
  \left|\smash{#3}{\vphantom1}^{+}\right\rangle}
\def\sandmm#1.#2.#3{%
  \left\langle\smash{#1}{\vphantom1}^{-}\right|{#2}%
  \left|\smash{#3}{\vphantom1}^{-}\right\rangle}
\def\sandpm#1.#2.#3{%
  \left\langle\smash{#1}{\vphantom1}^{+}\right|{#2}%
  \left|\smash{#3}{\vphantom1}^{-}\right\rangle}
\def\sandmp#1.#2.#3{%
  \left\langle\smash{#1}{\vphantom1}^{-}\right|{#2}%
  \left|\smash{#3}{\vphantom1}^{+}\right\rangle}

\def\Aloop{A^{\rm 1-loop}}

\def\Mloop{M^{\rm 1-loop}}

\def\Aloop{A^{\rm 1-loop}}

\def\NeqEight{{\cal N} = 8}
\def\NeqFour{{\cal N} = 4}

\def\BRQ#1#2#3{\la #1|{#2}|#3\ra}
\def\BBRQ#1#2#3{[ #1|{#2}|#3\ra}

\def\small{}

\begin{document}

\begin{titlepage}

\begin{flushright}

SWAT-06/481 \\

\end{flushright}

\vskip 2.cm

\begin{center}
\begin{Large}
{\bf SIMILARITIES OF GAUGE AND GRAVITY AMPLITUDES
}

\vskip 2.cm

\end{Large}

\vskip 2.cm
{\large
N.~E.~J.~Bjerrum-Bohr${}$,
David~C.~Dunbar${}$ and
Harald~Ita${}$
} 

\vskip 0.5cm

\vskip 0.5cm

{\it  Department of Physics \\
University of Wales Swansea \\
 Swansea, SA2 8PP, UK }

\vskip .3cm

\begin{abstract}

We review recent progress in computations of amplitudes in 
gauge theory and gravity. We compare the perturbative 
expansion of amplitudes in $\NeqFour$ super Yang-Mills and $\NeqEight$ 
supergravity and discuss surprising similarities.

\end{abstract}

\end{center}

\vfill

\end{titlepage}

 \section{Introduction}\label{sec1}
Perturbative gauge theory and gravity in four dimensions are quite dissimilar 
from a dynamical viewpoint. Gauge theory ({\it e.g.} pure Yang-Mills theory) is a 
renormalisable theory that is strongly coupled in the infrared and asymptotically 
free in the ultraviolet. Gravity on the other hand is a weakly coupled theory in 
the infrared but strongly coupled in the ultraviolet. 
By power counting, gravity in four dimensions is potentially a non-renormalisable 
theory. Pure gravity scattering amplitudes are finite at one-loop with the first 
divergence occurring at two-loops\,\cite{'tHooft:1974bx}.

Supersymmetry generally softens the UV behaviour in a quantum field theory. 
For example, maximally supersymmetric Yang-Mills is a finite theory\,\cite{Mandelstam}
and supergravity theories have a finite S-matrix until at least three 
loops\,\cite{Grisaru:1976ua}.
Although four-dimensional power counting and counter-term arguments suggest that supergravity 
theories are non-renormalisable\,\cite{Deser:1977nt} this has, so far, not been tested 
by direct computations. Arguments based on power counting within unitary cuts suggest 
that the first counter term in maximal supergravity\,\cite{ExtendedSugra} is expected 
at five loops\,\cite{BDDPR,Howe:2002ui}. 

Recently, initiated by the duality between gauge theories and a twister string 
theory\,\cite{Witten:2003nn}, there has been much progress in the computation of 
amplitudes in gauge theory. In this talk we discuss how these ideas may be applied 
to gravity calculations and the results thereof. 
We will first review the recent progress in computing physical on-shell tree amplitudes 
for gravity theories particularly focusing on the on-shell recursion 
relations\,\cite{Britto:2004ap,BCFWgravity} and the MHV-vertex construction\,\cite{CSW,CSWLoop,CSW:matter,CSW:massive,gravCSW}.  
Later we will discuss one-loop amplitudes. 
A surprising result is that the one-loop amplitudes of ${\cal N}=4$ SYM
and ${\cal N}=8$ supergravity\,\cite{BeBbDu,BBDuIt} occur to be expressible 
in terms of scalar box integral functions - despite the expectation from power 
counting. 
Supergravity multi-loop amplitudes are not directly addressed, however, 
the structure of amplitudes at tree-level and one-loop have, through factorisation 
and unitarity, important consequences on the structure of higher loop amplitudes.
\vskip -0.5 truecm
 \section{Old and New Techniques for Gravity Tree Amplitudes}

Graviton scattering amplitudes are extremely difficult to evaluate using conventional 
Feynman diagram techniques. In this section we review alternative methods: 1) the 
Kawai-Lewellen-Tye relations, 2) on shell recursion relations and 3) MHV vertex 
constructions.

{\bf 1)} Gravity amplitudes can be constructed through the Kawai, Lewellen 
and Tye (KLT)-relations\,\cite{KLT} as squares of gauge theory amplitudes.
The KLT relations are inspired by the na\"ive string theory relation
\vspace{-0.1cm}
$$
\hbox{closed string} \sim \hbox{(left-mover)} \times
                          \hbox{(right-mover)} \,,
\equn\label{StringRelation}
\vspace{-0.2cm}
$$
and have the explicit form, 
up to five points,
\begin{eqnarray*}
M_3^{\rm tree}(1,2,3) &=&
-iA_3^{\rm tree}(1,2,3)A_3^{\rm tree}(1,2,3)\,,\\
M_4^{\rm tree}(1,2,3,4) &=&
-is_{12}A_4^{\rm tree}(1,2,3,4)A_4^{\rm tree}(1,2,4,3)\,, \label{KLTFour} \\
M_5^{\rm tree}(1,2,3,4,5) &=& is_{12}s_{34}\ A_5^{\rm tree}(1,2,3,4,5)A_5^{\rm tree}(2,1,4,3,5) \\
&&  \null
\hspace{-0.57cm}\;+\;i s_{13}s_{24}\ A_5^{\rm tree}(1,3,2,4,5)A_5^{\rm tree}(3,1,4,2,5)\,,\label{KLTFive} 
\vspace{-0.3cm}
\end{eqnarray*}
where
$A^{\rm tree}_n$ are the tree-level colour-ordered gauge theory partial amplitudes. 
We suppress factors of $g^{n-2}$ in the $A^{\rm tree}_n$ and
$(\kappa/2)^{n-2}$ in the $M_3^{\rm tree}$. 

The KLT relations are helpful in the calculation of gravity tree amplitudes, 
however they have some undesirable features. The factorisation structure is not 
manifest and the expressions do not tend to be compact, as the permutation sums
grow rather quickly with $n$. 
In fact, the Berends, Giele and Kuijf (BGK) form of the MHV gravity 
amplitude\,\cite{BerGiKu},
$$ 
\eqalign{
M_n^{\rm tree}
(1^-,2^-,3^+,\cdots, n^+)=-i\spa1.2^8\times
\biggl[\biggl({ \spb1.2\spb{n-2}.{n-1}    \over \spa1.{n-1} N(n) }\biggl)\times
\cr
\times\biggl(  \prod_{i=1}^{n-3} \prod_{j=i+2}^{n-1} \spa{i}.j \biggl)
\prod_{l=3}^{n-3} (-[n|K_{l+1,n-1}|l\ra)
+{\cal P}(2,3,\cdots,n-2)
\biggr]\,,
\cr}
$$is rather more compact than that of the KLT sum (as is the expression
in \cite{BCFWgravity}.)
In the above we use the definitions,
$\BBRQ k {{K}_{i,j}} l \;\equiv\; \BRQ {k^+} {\Slash{K}_{i,j}} {l^+} \;\equiv\; \BRQ
{l^-} {\Slash{K}_{i,j}} {k^-} \;\equiv\; \la l |{{K}_{i,j}}| k] \;\equiv\; \sum_{a=i}^j\spb k.a\spa a.l\,,$ 
and $N(n)=\prod_{1\leq i<j \leq n} \spa{i}.{j}$.
In terms of the above Weyl spinors we often use twistor variables
$\lambda^a_i \equiv | k^+_i \ra$ and $\bar\lambda^{\dot a}_i \equiv | k^-_i \ra$.
The MHV amplitudes for graviton scattering display a feature not
shared by the Yang-Mills expressions, they depend not only on the
holomorphic variables $\lambda$, but also on the anti-holomorphic
$\bar\lambda$ variables (within the $s_{ij}$ for the KLT
expression). 

{\bf 2)}
In a recent computational approach for amplitudes, Britto,
Cachazo, Feng and Witten\,\cite{Britto:2004ap} obtained on-shell recursion 
relations for trees. The recursion relations are based on factorisation 
properties of amplitudes and are thus applicable to a wide range of theories and 
in particular to gravity\,\cite{Britto:2004ap,BCFWgravity}.
The technique is based on analytically shifting a pair of external legs,
$
\lambda_i \;\rightarrow\;  \lambda_i \;+\;z \lambda_j,
\bar \lambda_j \;\rightarrow\;  \bar\lambda_j \;-\;z  \bar\lambda_i\,,
$ 
and on determining the physical amplitude, $M_n(0)$, from the poles in the
shifted amplitude, $M_n(z)$.
This leads to a recursion relation of the form,
\begin{eqnarray*}
M_n(0) \;=\;  \sum_\alpha  \hat M_{n-k_\alpha+2}(z_\alpha)\times {
i \over P_\alpha^2}
         \times \hat M_{k_\alpha}(z_\alpha)\,,
\end{eqnarray*} 
where the factorisation is only on these poles, $z_\alpha$,
where legs $i$ and $j$ are connected to different sub-amplitudes. 
An essential condition for the recursion relations is that the shifted amplitude $M_n(z)$ 
vanishes for large $z$. Whereas proven for gauge theory amplitudes a general proof 
(for arbitrary helicities\,\cite{BCFWgravity}) in gravity is an open problem. 

Recursion relations based on the analyticity in the complex plane can also be used at 
loop level both to calculate rational terms\,\cite{OneLoopQCD} and the coefficients of 
integral functions\,\cite{Bern:2005hh}.

{\bf 3)}
Finally, we would like to mention the CSW construction\,\cite{CSW,Kasper} of amplitudes and 
its generalisation to gravity\,\cite{gravCSW}. In this approach MHV-amplitudes are treated 
as fundamental vertices and generic scattering amplitudes are expanded in terms of these 
MHV-vertices. 

Considering the N$^s$MHV amplitude with $n$ external legs. One would begin by drawing all 
diagrams which may be constructed using MHV vertices. 
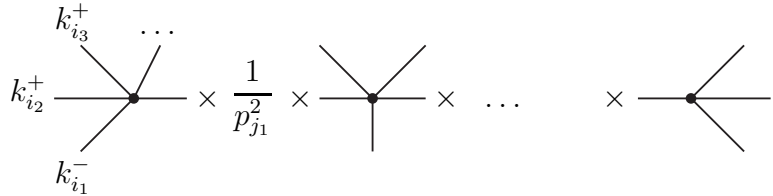
\vspace{1cm}
\begin{center}
\begin{picture}(100,75)(80,-40)
\SetWidth{0.7}
\Line(30,30)(0,30)\Line(30,30)(10,50)\Line(30,30)(40,50)\Line(30,30)(10,10)
\Line(30,30)(50,30)\Line(120,30)(100,50)\Line(120,30)(140,50)
\Line(120,30)(100,30)\Line(120,30)(140,30)\Line(120,30)(120,10)
\Line(240,30)(220,30)\Line(240,30)(260,50)\Line(240,30)(260,10)
\Line(240,30)(270,30)
\Vertex(30,30){2}\Vertex(120,30){2}\Vertex(240,30){2}
\Text(7,1)[c]{$k_{i_1}^-$}\Text(-10,32)[c]{$k_{i_2}^+$}
\Text(7,59)[c]{$k_{i_3}^+$}\Text(39,55)[c]{$\ldots$}
\Text(58,30)[c]{$\times$}\Text(93,30)[c]{$\times$}
\Text(75,30)[c]{$\displaystyle\frac{1}{p_{j_1}^2}$}
\Text(160,30)[c]{$\times\ \ \ldots$}\Text(212,30)[c]{$\times$}
\end{picture}
\vspace{-1.4cm}
\end{center}
The contribution from each diagram is a product of $(s+1)$ MHV vertices and $s$
propagators as indicated above.

\noindent The contribution of a given diagram to the total amplitude can be calculated by
evaluating the product of MHV amplitudes and propagators,
\begin{eqnarray*}
M^s_N{\big |}_{\mbox{\small CSW-diagram}}\;=\;\left(\prod_{l=1,s+1} M^{{\rm MHV}}_{N_l}(\hat K_l)\right)
\prod_{j=1,s} \frac{i}{ p_{j}^2}\, ,
\label{CSWrule}
\end{eqnarray*}
where the propagators are computed on the set of momenta $k_i$ and
$p_j$, and the MHV vertices are evaluated at shifted momenta $\hat
k_i$ and $\hat p_j$. The momenta $k_i$ are external and the momenta $p_{j}$ 
internal, and given by momentum conservation at each MHV-vertex. 
A key feature is the interpretation of the MHV amplitudes for internal legs. 
For Yang-Mills where the MHV vertices only depend on $\lambda$ 
the correct interpretation is\,\cite{CSW}
$$
\lambda(p)_a = p_{a\dot a} \eta ^{\dot a}\,, 
$$ for an arbitrary reference spinor $\eta$. For gravity amplitudes we
must also solve for $\bar\lambda(p)$ which is
less obvious\,\cite{GBgravity} and will be a function of the momentum of the 
negative helicity legs. It turns out that all spinors are uniquely defined in terms of 
the shifted momenta $\hat k_i$ and $\hat p_{j}$ if we demand that 
they are null vectors obeying momentum conservation at each vertex.

Explicitly they are given by shifting the negative helicity legs $\bar\lambda_i$
by
$$
\bar\lambda_i \longrightarrow \bar\lambda_i +a_i \bar\eta\,, 
$$
and leaving $\bar\lambda_j$ of the positive helicity legs $k_{j^+}$ untouched. 
The 
$s+2$ parameters, $a_{i^-}$
are uniquely fixed\,\cite{gravCSW} by demanding a) overall momentum conservation, 
b) momentum conservation at each vertex and finally c) that the internal 
momenta, $\hat p_{j}$, are massless $\hat p_{j}^2=0$.

As an example of how the MHV-vertex constructions works for gravity,
we can consider the $n$-point NMHV amplitude with three
negative helicity legs $m_i$.  The MHV-vertex approach gives the
amplitude in the form,
$$
\sum_{\rm perms,{\it r}} 
M^{{\rm MHV}}( \hat k_1^-, k_4^+,\cdots k_r^+, \hat p^- )
\times { 1 \over p^2 } 
\times 
M^{{\rm MHV}}( \hat k_2^-, \hat k_3^- k_{r+1}^+, \cdots k_{n-3}^+, \hat p^+ )\,,
$$ where the sum runs over diagrams involving all choices of $r> 0$ and all
permutations of the negative and positive helicity legs. To illustrate the
correct continuation, the three negative helicity $\bar\lambda$ must
be shifted $\bar\lambda_i\rightarrow\bar\lambda_i+a_i\bar\eta$.
Imposing the momentum constraints leaves us with a shift,
$$
\bar\lambda_1 \longrightarrow \bar\lambda_1 +z\la k_2, k_3 \ra \bar\eta\,,
$$
together with the cyclic shifts of the other two legs. Momentum is
conserved for any value of the parameter $z$. Requiring $\hat p^2=0$ then fixes $z$
uniquely as $z = { p^2 / [\eta | p | k_1 \ra}$ and the MHV vertex 
expansion is completely determined. 

\section{One-Loop Amplitudes in $N=8$ Supergravity}
In a Yang-Mills theory, the loop 
momentum polynomial in a one-loop $n$-point diagram will generically be of degree $\leq n$.
$\NeqFour$ one-loop amplitudes exhibits 
considerable simplification and the loop momentum 
integral will be of degree $n-4$\,\cite{StringBased,BDDKa}. 
Consequently,  from a Passarino-Veltman 
reduction\,\cite{PassVelt}, the amplitudes can be expressed as a sum of scalar box 
integrals with rational coefficients, 
\vspace{-0.1cm}
$$
\Aloop= \sum_{a}  c_a I^4_a\,.\equn\label{OnlyBoxesEQ}
\vspace{-0.2cm}
$$
Determining the amplitude then reduces to determining the rational coefficients $c_a$.
Inspired by the duality in \cite{Witten:2003nn}, considerable progress has recently been 
made in determining such coefficients, $c_a$, using a variety of methods 
based on unitarity\,\cite{BDDKa,BDDKb,NeqFour,NeqOne}. 

For $\NeqEight$ supergravity the equivalent power counting 
arguments\,\cite{GravityStringBased} give a loop momentum 
polynomial of degree
$$
2(n-4)\,,
$$ 
which is consistent with Eq.~(\ref{StringRelation}).  
Reduction for $n>4$ leads to a sum of tensor box integrals 
with integrands of degree $n-4$ which would then reduce
to scalar boxes {\it and} triangle, bubble and rational 
functions,
$$
\Mloop= \sum_a c_a I_4^a +\sum_a d_a I^a_3 +\sum_a e_a I^a_2 +R\,,
$$
where the $I_3$ are present for $n\geq 5$, $I_2$ for $n \geq 6$ and $R$ for $n\geq 7$. 

There is evidence that all one-loop amplitudes of $\NeqEight$, 
like the $\NeqFour$ amplitudes Eq.~(\ref{OnlyBoxesEQ}), can
be expressed as a sum over scalar box integrals, the so called 
``no-triangle hypothesis''\,\cite{BeBbDu,BBDuIt}.
Firstly, in the few definite computations at one-loop level, 
triangle or bubble functions do not appear. The first computation was of the four-point
amplitude\,\cite{GSB} where only box functions appear (although this is consistent with power counting). 
Beyond this computations of the 
five and six point MHV-amplitudes 
yielded only
scalar box-functions\,\cite{Bern:1998sv}. 

Secondly, the factorisation properties of physical amplitudes 
do not demand the presence of these functions. Since the
four and five point amplitudes are triangle-free then in any
factorisation limit of a higher point function the triangles 
must vanish. 
In this spirit an ansatz for the $n$-point MHV amplitude 
was constructed\,\cite{Bern:1998sv} entirely of 
box functions consistent in all soft limits.

Thirdly, one can check whether the amplitudes composed purely from box-functions precisely give  
the expected soft divergence in a $n$
graviton amplitude\,\cite{Dunbar:1995ed},
\vspace{-0.1cm}
$$\small
\eqalign{
M^{\rm one-loop}_{
[1,2,\ldots, n]}
=
{i c_\Gamma \kappa^2}
\bigg[
{\sum_{i<j}  s_{ij} \ln[-s_{ij} ]
\over 2\epsilon}
\bigg]
\!\!\times\! M^{\rm tree}_{[1,2,\ldots, n]}.
\cr}
\label{IRamplitudeEQ}
\vspace{-0.15cm}$$ 
In~\cite{BBDuIt} the box coefficients were explicitly computed
for the six-point NMHV amplitudes confirming the above claims.  

The ``no-triangle'' hypothesis applies to one-loop amplitudes. However, by
factorisation it should have implications beyond one-loop suggesting
the UV behaviour of maximal supergravity may be significantly milder
than expected from power counting.

\end{document}